\documentclass[pre,aps,amsmath,amssymb,
 preprint,%
 reprint,%
 twocolumns,
 superscriptaddress,
 floatfix,
 showpacs,
]{revtex4-1}

\usepackage{graphicx}% Include figure files
\usepackage{bm} % bold math
\usepackage{epsfig}
\usepackage{color}

\usepackage{soul}
\usepackage{cancel}

\newcommand{\be}{\begin{equation}}
\newcommand{\ee}{\end{equation}}
\newcommand{\bea}{\begin{eqnarray}}
\newcommand{\eea}{\end{eqnarray}}
\newcommand{\ba}{\begin{array}}
\newcommand{\ea}{\end{array}}

\begin{document}

%\preprint{}
\title{Active translocation of a semiflexible polymer assisted by an
  ATP-based molecular motor}

\author{A. Fiasconaro}
\email{afiascon@unizar.es}

\affiliation{Dpto. de F\'{\i}sica de la Materia Condensada,
Universidad de Zaragoza. 50009 Zaragoza, Spain}

\author{J.J. Mazo}

\affiliation{Dpto. de F\'{\i}sica de la Materia Condensada,
Universidad de Zaragoza. 50009 Zaragoza, Spain}

\affiliation{Instituto de Ciencia de Materiales de Arag\'on,
C.S.I.C.-Universidad de Zaragoza. 50009 Zaragoza, Spain.}

\author{F. Falo}

\affiliation{Dpto. de F\'{\i}sica de la Materia Condensada,
Universidad de Zaragoza. 50009 Zaragoza, Spain}

\affiliation{Instituto de Biocomputaci\'on y F\'{\i}sica de Sistemas
Complejos, Universidad de Zaragoza. 50018 Zaragoza, Spain}

\date{\today}

\begin{abstract}
In this work we study the assisted translocation of a polymer across a membrane nanopore, inside which a molecular motor exerts a force fuelled by the hydrolysis of ATP molecules. In our model the motor switches to its active state for a fixed amount of time, while it waits for an ATP molecule binding and triggering the impulse, during an exponentially distributed time lapse. The polymer is modelled as a beads-springs chain with both excluded volume and bending contributions, and moves in a stochastic three dimensional environment modelled with a Langevin dynamics at fixed temperature. The resulting dynamics shows a Michaelis-Menten translocation velocity that depends on the chain flexibility. The scaling behavior of the mean translocation time with the polymer length for different bending values is also investigated.
\end{abstract}

\pacs{05.40.-a, 87.15.A-, 87.10.-e, 36.20.-r}

% 05.40.-a  Fluctuation phenomena, random processes, noise, and Brownian motion
% 87.15.A-  Theory, modelling, and computer simulation
% 87.10.-e  General theory and mathematical aspects
% 36.20.-r  Macromolecules and polymer molecules
% 87.15.H   Dynamics of biomolecules
% 87.23.Cc
%\PACS 05.40-a,87.23Cc,89.75-k, 87.17.Aa, 82.20.-w

%Statistical Mechanics, Mean First Passage Time, Noise-induced effects, NES
\keywords{Stochastic Modeling, Fluctuation phenomena, Polymer
dynamics, Langevin equation, Molecular simulation}

\maketitle

\section*{Introduction}

Translocation of long molecules through nanopores in cell membranes is a common process in living systems. Cell drug delivery, DNA, RNA and
protein passage through cell membranes and nuclear pores, and DNA
injection and packaging by phage viruses are only some interesting
examples of a broad biological phenomenology.~\cite{MetzlerSM2014}

The passage of polymers thorough nanopores is also a
fundamental problem in chemical and industrial processes. In this context, many efforts are made in nanotechnological applications, that try to emulate the complex biological processes involved in the
translocation problem~\cite{li2001,mickler,starikov}. An important related application is the use of nanopores to unzip and translocate single DNA chains with the purpose of perform fast and detailed DNA
sequencing~\cite{sigalov2008,Merchant2010,Fyta2011,SciRep2016,GolestPRX}.

In spite of the many studies present in the literature, the
understanding of polymer translocation at the nanoscale still deserve deeper investigation. In fact, the number of parameters involved in the models and the presence of nonlinearities together with the underlying stochastic environment in which nanoscale objects move, make this study complex theoretically and hard computationally, even nowadays. DNA chains are paradigmatic examples of this difficulty. With the ambition of
efficiently describing the biological matter in a reasonable time,
different mesoscopic models for polymer translocation have been
introduced~\cite{Meller2003}. In some very simplified models a single
barrier potential, eventually depending on time, is introduced to
depict the overall translocation
process~\cite{Pizz2010,Pizz2013,Sung1996,Muthukumar}; in others,
stochastic and ratchet-like forces and potentials are
used~\cite{linke,linke2,shulten}.

\begin{figure}[tb]
\centering
\includegraphics[width=8.5cm]{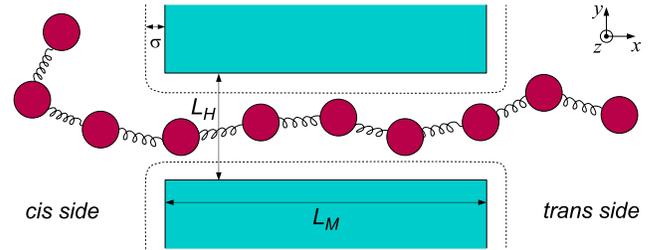}
\caption{Section of the polymer translocating through a nanopore
in 3d. The pore  has a square section of width $L_H$ and its
length is $L_M$. Inside the pore it is present a driving force in
the $x$-direction, which pushes the polymer toward the
\emph{trans} side of the membrane. The walls of the membrane
repulse uniformly the chain  inside a characteristic distance
$\sigma$.} \label{schema}
\end{figure}

Motivated by the passive pores studied in different
experiments~\cite{KasPNAS96,RMP,NL}, most studies of translocation
have been performed, so far, mainly under constant forces. These forces, are generally associated to potential and concentration gradients observed
between both sides of the membrane. The role of active nanopores,
with mechanisms to assist the polymer translocation, has been only
more recently considered. The simplest active nanopore corresponds to
the opening or closing of the channel due to either the presence of
electrochemical membrane forces~\cite{2002Alberts} or random bonding
of ligands~\cite{1998Petersen}. It has been usually modelled by a
dichotomous Markov noise (also called random telegraph
noise)~\cite{ajf-rtn,ikonen2012,ajf-sin-rtn-3D} which drives the
molecule. In other cases, the flickering activity of the pore channel
is modelled as a sinusoidal pore
actuation{~\cite{ajf-sin,golest2011,ikonen2012,golest2012}}. In both
cases a new time scale, associated to the pore activity, appears in the description of the phenomena.

A landmark example of active translocation in the biological realm is the translocation assisted by molecular motors.  Recent well known
experiments have risen a considerable attention to the specific and
very interesting driving features of the molecular motor of the virus
bacteriophage $\phi$-29. This motor is able to inject out (or pack
inside) its DNA by exerting a force supplied by the hydrolysis
of ATP molecules which bind to the motor sites and activate its
movement~\cite{Bust01,Bust09}. The force exerted by the nanomotor can
be modelled by a specific stochastic dichotomous force, that we call
dichotomous ATP-based motor noise force. This force has been
successfully introduced to study translocation restricted to a one
dimensional dynamics.~\cite{ajf-damn,pffs-damn}

In addition to its biological interest, understanding and controlling
bistable forces has an undoubted interest in the manipulation of
objects at the nanoscale. In fact, molecular motors represent a highly effective way to exert a force at the nanoscale worth to be mimicked in the artificial realm. So, a properly prepared system can be driven in a controlled way by changing the concentration of the fuel molecules in the solution, which directly modifies the motor activation rate, and thus the exerted force.

Concerning the polymer modelling, the usual approach is based in
the Rouse chain~\cite{Rouse} and its modifications. There, polymers
are constituted by beads connected each other by harmonic
springs. Natural improvements of the model take into account polymer
bending energy and excluded volume interactions. Indeed, recent
studies show that translocation significantly depends not only on the
polymer size but also on its flexibility and on the interactions
between monomers and
pores~\cite{sigalov2008,slater2013,larrea2013,ikonen2012-2,ajf-sin-rtn-3D}.

Aimed by all these results, this manuscript proposes the study of the
translocation of a linear polymer molecule in the 3d domain driven by
a specific ATP-fuelled biological force acting inside the nanopore. We will specially focus on the study of the mean translocation velocity and translocation time
(TT) $\tau$ of the polymer as a function of the motor activation
rate. We will put special attention to the study of the dynamics of
the system for different polymer size (number of monomers) and
persistence length (flexibility or bending). This latter is an important parameter of the system whose influence has been rarely studied so far in translocation context.

\begin{figure}[t]
\centering
\includegraphics[angle=-90,width=8.5cm]{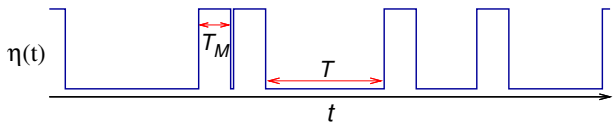}
\caption{Scheme of the dichotomous pushing force acting on the polymer. $T_M$ is the working time of the
motor, supposed fixed. $T$ is the mean waiting time in the inactive
state. The motor acts on a region of length $L_M$, see Fig.~\ref{schema}.}
\label{DAMN}
\end{figure}

\section*{The model}

\subsection*{System equations}

Our polymer model is based in a chain formed by $N$ identical monomers moving in the three-dimensional space.  We use a modified Rouse
model~\cite{Rouse} which includes bending energy, excluded volume
effects and interaction with the membrane and the pore. The elastic
potential energy is given by
 \be V_{\rm el}(d_i)=\frac{k_e}{2}\sum_{i=1}^{N} (d_i-d_0)^2,
 \label{v-har}
 \ee
\noindent where $k_e$ is the elastic parameter, $\bm{r}_i$ is the
position of the $i$-th particle,
$d_i=|\bm{d}_i|=|\bm{r}_{i+1}-\bm{r}_i|$, is the distance between the
monomers $i$ and $i+1$, and $d_0$ is the equilibrium distance between
adjacent monomers.

The model takes into account the bending energy of the chain with a term given by
 \be
  V_{\rm ben}(\theta_i)=\frac{k_b}{2}\sum_{i=1}^{N} [1-\cos(\theta_i-\theta_0)],
 \label{v-ben}
 \ee
where $k_b$ is the bending elastic constant, $\theta_i$ is the
angle between the links $\bm{d}_{i+1}$ and $\bm{d}_{i}$, and $\theta_0$ the equilibrium angle, with $\theta_0=0$ in our case. With this term, our model is a discrete version of the worm-like chain (WLC) model.

In order to consider excluded volume effects between the monomers, a repulsive only Lennard-Jones potential has been taken into account:
 \be V_{\rm
   LJ}(r_{ij}) = 4 \epsilon \sum_{i \ne j=1}^{N} \left[ \left(\frac{\sigma}{r_{ij}}\right)^{12}-
   \left(\frac{\sigma}{r_{ij}}\right)^6 \right]
 \label{LJ}
 \ee
for $r_{ij}\leq 2^{1/6}\sigma$, and $-\epsilon$ otherwise. Here $r_{ij}$ is the distance between monomer $i$ and monomer $j$.

The dynamics of every monomer of the chain is obtained by the
overdamped equation of motion
 \bea
  m\gamma\dot{\bm{r}_i} = &-& \nabla_i V_{\rm el}(d_i)
-\nabla_i V_{\rm ben}(\theta_i) - \nabla_i V_{\rm
LJ}(r_{ij}) \nonumber \\
 &+& F_{drv,i} \bm{i} + \bm{F}_{sp,i} + \sqrt{2m\gamma k_BT} \, \bm{\xi}_i(t),
 \label{eq}
 \eea
where the effective viscosity parameter of each monomer is
included in the normalised time units. $\bm{\xi}_{i}(t)$ stands
for the Gaussian uncorrelated thermal fluctuation and follows the
usual statistical properties $\langle\xi_{i,\alpha}(t)\rangle=0$
and $\langle\xi_{i,\alpha}(t)\xi_{j,\beta}(t')\rangle = \delta_{i
j}\delta_{\alpha,\beta}\delta(t'-t)$, with $i=1,...,N$, and
$\alpha = x,y,z$. The operator $\mathbf{\nabla}_i =
\partial / \partial x_i \, \bm{i} + \partial / \partial y_i
\, \bm{j}  + \partial / \partial z_i \, \bm{k}$.

The force term $\bm{F}_{sp}$ includes both the chain-membrane and
chain-pore spatial constraint. This interaction force is modelled with
the same repulsive Lennard-Jones potential described in
Eq.~(\ref{LJ}), also with the same parameter values. It takes place
uniformly and perpendicularly to all the planes that define both the
membrane and the pore channel, modelled as a square prism of base $L_M$ and length $L_H$ (see Fig.~\ref{schema}).

Finally, $F_{drv,i}$ is the driving force which represent the force
$F_{drv}$ acting on particle $i$ when it is inside the pore. $F_{drv}$ is constituted by two terms: a constant force $F_c$
and the motor force $\eta(t)$. $F_c$ is the constant force used in the study of passive pores (usually associated to potential and
concentration gradients between both sides of the membrane). $\eta(t)$ is associated to the molecular motor pulling the polymer. It
fluctuates between 0 and $F_M$. For $\eta(t)$ we use a modification of the usual dichotomous noise~\cite{ajf-damn}. Activated by ATP absorption, $\eta$ acts during a fixed time $T_M$, along this time $\eta(t)=F_M$. After this time, the motor relax and $\eta(t)=0$ until a new ATP molecule is absorbed. This happens with a probability $P_{t'}=1-{\rm e}^{-t'/T}$, being $t'$ the time spent by the motor in its inactive state, and $T$ its mean {\em waiting time} (see Fig.~\ref{DAMN}). We define $\nu = 1/T$ as the related \emph{activation rate} which is assumed to be proportional to
the ATP molecule concentration.

The motor acts only on the monomers inside the pore. Thus,
regarding the spatial dependence of the total driving force:
 \be
   F_{drv}(x) = \left\{
    \begin{array} {lr}
        (F_c, F_c+F_M)   & x \in [0,L_M] \vspace{0.2cm}
    \\
         0          & otherwise.
    \end{array} \right.
   \label{Spatial}
 \ee

\subsection*{ATP Energy supply and Michaelis-Menten kinetics}

The energy used by the motor to provide the driving is given by the
ATP hydrolysis.  The simplest way to model this phenomenon, is to
consider that each ATP hydrolysis activates the motor for a fixed
working time $T_M$. The ATP binding and subsequent absorption is
itself a Poisson process which depends on the ATP concentration in
the solution surrounding the motor. Once absorbed ATP is hydrolysed
releasing energy, and the motor changes its conformational state and
exerts a force $F_M$ (considered constant here) during a fixed time
$T_M=1/\nu_0$. After that time the motor returns to its rest state. A
new force will be applied when the next ATP suitable quantity is
absorbed and hydrolysed, which happens after a mean time $T=1/\nu$,
which follows an exponential distribution of waiting times according
to the underlying Poisson process.  The timing in the kinetics can be
summarised as follows: For activation rates $\nu < \nu_0$, the motor
remains at rest for an average time longer than $T_M$, while the
opposite occurs (average rest-state time of the motor smaller
than $T_M$) if $\nu > \nu_0$.

The kinetic feature of the motor, i.e. that it works for a fixed time, and that the statistics of the arrival of the ATP molecules happens in an exponential distribution, is a good realistic approximation as put in evidence by different experimental works~\cite{Bust01,Bust05,Bust09}.
With the above definitions, the Michaelis-Menten (MM) law for the transfer velocity arises naturally from the model. In fact, the fraction of active time with respect to the total time is given by
 \be
  \frac{T_M}{T_M + T}=\frac{\nu}{\nu_0+\nu} = \frac{[ATP]}{k_M+[ATP]},
  \label{MM}
 \ee
which is a MM law. The only hypothesis included in
the derivation of the last equality is that the motor activation rate
$\nu$ is proportional to the ATP concentration ($\nu \propto [ATP]$),
being $k_M=[ATP]_0$, the ATP concentration at $\nu = \nu_0$. In our approach the binding of an ATP molecule to the motor is a Poisson process, with a binding rate that can be considered, in a first approximation, proportional to the ATP concentration. This reasonable assumption is compatible with the experimental observation reported in~\cite{Bust09}. There, data are adjusted to a more complex functional relation, but it is easy to show that the used fitting law also results in a Michaelis-Menten velocity after a suitable re-scaling of the involved parameters (see also~\cite{sancho}).

The relationship between the mechanical active-rest kinetics of the
motor and the MM law goes beyond the statistical ansatz
$\nu \propto [ATP]$, and represents a general paradigmatic behaviour
connecting chemical reactions with mechanical transfer of energy in
the context of microscopic description of single molecules dynamics.
In fact, it is possible to depict the ATP hydrolysis in the
context of the MM enzymatic reaction scheme:
 \be E + S
\longrightarrow \!\!\!\!\!\!\!\!\! ^{k_1} \ \ Z \longrightarrow
\!\!\!\!\!\!\!\!\!  ^{k_2} \ \ E + P,
 \ee
where the rate $k_1$ represents the probability to form the compound $Z$ per unit of time and per unit of $S$ ([ATP]), and $k_2$ gives the probability to form the product $P$ per unit of time. In our mechanical and individual case (single motor and single ATP hydrolisation event) $Z$ represents the ATP-motor binding, which occurs with the rate $\nu$ (\emph{i.e.} $\sim k_1 [ATP]$), while the product of the reaction $P$ represents the motor action which is completed within a time $T_M =1/\nu_0$ ($\nu_0 \sim k_2$). These relationships are in agreement with the definition of the Michaelis constant $k_M=k_2/k_1$. For a more complete analysis of the MM law the reader can also refer to Ref.~\cite{engl,mof}.

\subsection*{Polymer velocity}
The polymer velocity (along the $x$-axis) is calculated by summing up
the $N$ monomer terms of equation~(\ref{eq}) and averaging over the
time of the translocation. The mean velocity of the centre of mass of
the chain is then:
 \bea
   v_{CM} &=&  \frac{F_{drv}}{N} + \frac{F_{w}}{N} = \frac{F_c+F_{M}}{N}\sum_{i=1}^N \langle \eta_i(t) \rangle + \frac{F_{w}}{N} =
              \nonumber \\
          &=& \frac{F_c n_{\rm M}}{N} + \frac{F_M}{N} \frac{n_{\rm M}}{1+\nu_0/\nu} + \frac{F_{w}}{N}.
 \label{mmeq}
 \eea
In the above equation $n_{\rm M}$ is the mean number of
monomers inside the motor during the translocation, a number which
weakly depends on $\nu$, $k_b$ and $N$, and we consider here as a
constant, $n_M=5$, as confirmed by our numerical observations.

We identify here three contributions to the total velocity. The first term accounts for the driving of the constant external force, which acts only on the monomers into the motor. The second term is related with the molecular motor kinetics, explained above. It is given by the force felt by the $n_{\rm M}$ monomers inside the machine which
operates for the fraction of time $\nu/(\nu_0 + \nu)$, and shows the
Michaelis-Menten dependence. The last term $F_{w}(\nu, k_b, N)$ comes from the interaction of the polymer with the membrane. In fact, while all the internal force terms in equation~(\ref{eq}) sum up to zero when we sum the $N$ equations of the monomers, the interaction with the walls in the $x$-direction does not average to zero, being a non symmetric force reaction, as the pulling goes to positive $x$-direction. The resulting force opposes to the righthand movement of the chain, and depends on the activation rate of the motor $\nu$, the length $N$, and the rigidity of the chain $k_b$. It is important to highlight that all the $k_b$ dependence in the polymer velocity comes from this term. We will also see below that although $F_w$ also depends on $\nu$, the overall $v_{CM}(\nu)$ dependence follows the MM behaviour, dominated by the motor contribution to the total force.
Note also that the polymer velocity goes to zero as $1/N$ for large chains as expected when a motor acts on a small number of monomers $n_M$ of a polymer which moves in a dissipative media.

\section*{Numerical simulations}

\subsection*{Definitions and simulation details}

We are interested in characterising the translocation process of the
polymer through the pore. We will present below the results for
$\tau$, the mean value of the TT of the chain with different
parameters of the model and different polymer lengths.

It is usual in literature that a channel with either zero or short
length is used to study the polymer translocation. In order to
approach a more realistic picture, in our works we use a channel with
a fixed length, $L_M=5 d_0$, longer than the distance of two
consecutive monomers. We calculate the translocation time as follows:
All the simulations start with five monomers at the right end of the
polymer lying inside the pore and the others linearly ordered and
sited at the equilibrium distance. During a thermalisation time $t_t
= 1000$t.u. the chain evolves under the action of thermal fluctuations while keeping fixed the position of the five monomers inside the pore. After that transient time, the restriction over the first five
monomers is removed, and the evolution of the chain under the dynamics given by Eqs.~(\ref{eq}) is monitored. Sometimes the polymer moves
backwards into the {\em cis} region of the system, leaves the pore and it does not translocate. These cases are not taken into account,
\emph{i.e.} they do not enter in the translocation statistics. On the
contrary, whenever the polymer crosses the membrane the simulation
ends when the last monomer of the chain enters the pore, and the time
lasted from the end of the thermalisation process to the entrance of
the last monomer into the pore defines the translocation time (TT) of the event. Later we will average over many translocation processes to get the values of $\tau$.

At the same translocation conditions, we define the center of mass
(CM) velocity of the polymer $v_{CM}$ as the ensemble average (over
$N_{exp}$ realisations) of the ratio between the displacement of
the CM of the chain and the corresponding employed time (the
translocation time in each realisation).  It is worth to remember that the relation between translocation velocity $v_{CM}$ and mean time
$\tau$ is not immediate since in general, $\langle 1/t \rangle \neq
1/\langle t \rangle$.

The given definitions are suitable to compare TT of chains with different lengths because the border effects result minimised since five monomers are always inside the pore during the simulation time in all the cases studied.

Following \cite{ikonen2012}, we define $m$, $l_0$, and $\epsilon_0$ as the mass, the length, and the energy unit units respectively. This choice determines a Lennard-Jones time scale
given by $t_{LJ}= (ml_0^2/\epsilon_0)^{1/2}$. However, as the
dynamics we propose is overdamped, the time scale that normalise the
equation of motion Eq.~(\ref{eq}) is $t_{OD} = \gamma {t_{LJ}}^2 $,
thus depending on the damping parameter. To set some values, let us
consider a DNA molecule at room temperature ($k_BT= 4.1\,\rm{pN nm}$)
and the simplest model with $k_b=0$. We have fixed our simulation
temperature to $k_BT=0.1$ in dimensionless units. This choice fixes our energy unit in $\epsilon_0= 41\,\rm{pN nm}$. By setting $l_0 = 1.875\,\rm{nm}$ and $m = 936\,\rm{amu}$ \cite{ikonen2012}, we obtain $t_{LJ} \approx 0.38$ps, while the force unit is given by $\epsilon_0/l_0 = 21.9\,\rm{pN}$. An
estimation for the kinetic damping is $\gamma \approx 1.6 \times
10^{13}\,\rm{s^{-1}}$, so obtaining $t_{OD} \approx 2.3\,\rm{ps}$. Other normalisations can be used depending on the system to
simulate~\cite{af2012}.

The dimensionless geometrical values used in the simulations are
$L_H=2$, $L_M=5$. The rest distance between adjacent monomers is
$d_0=1$ and $k_e = 1600$, large enough to maintain the
bonds of the chain rigid enough. The Lennard-Jones energy is $\epsilon=0.3$, and $\sigma=0.8$. The values of $d_0$, $\sigma$, $L_M$ and $L_H$ guarantees that the polymer is maintained almost linear and ordered inside the pore. Also, the different choices of the bending constant $k_b$, gives the possibility to study the TT for different \emph{persistence lengths} of the chain, which basically gives the stiffness of the polymer (\emph{i.e.} its resistance to bend). For our model $L_p=k_b/k_BT$. Thus for example we obtain $L_p=5$ (in units of $d_0$) for $k_b=0.5$. Regarding the actuating forces we set $F_c=0.1$ and $F_M=0.2$ along our work.

For every set of parameters, a number of numerical experiments $N_g$
has been simulated. From them a number of $N_{exp}=2000$ successful
translocations have been reached and both the mean translocation time
$\tau$ and the mean velocity $v_{CM}$ have been then computed. From these values, we also compute the translocation
probability $P_{in}$ as the ratio $N_{exp}/N_g$. To finish we have to
mention that since we deal with time dependent pulling forces inside
the pore, we have also averaged on the initial value of the driving by randomising the initial phase of the force, that is considered as a
random variable with a uniform distribution over its possible values.

\subsection*{Polymer velocity}

\begin{figure}[t]
\centering
\includegraphics[angle=-90, width=8.5cm]{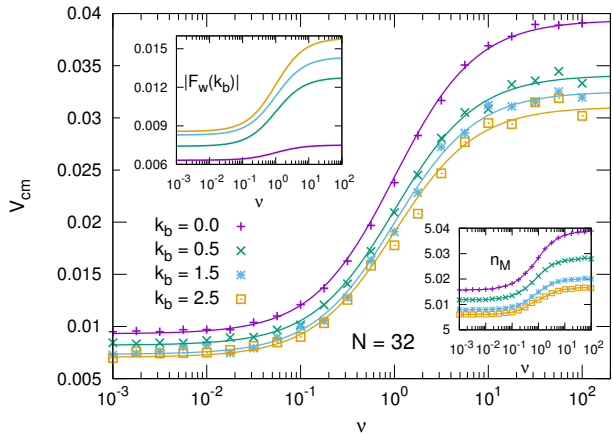}
\caption{Mean velocity of the polymer center of mass with $N=32$ for
 different values of the bending $k_b = 0.0, 0.5, 1.5, 2.5$. Symbols stand for numerical simulations, full lines show the MM best fit according to Eq.~(\ref{eqv}). The right inset shows the corresponding mean number of monomers inside the motor $n_M$ during the simulations.  Left inset: absolute value of the mean reaction force exerted by the walls during the translocation as a function of the motor actuation rate $\nu$.}
 \label{Fig:vcmMM}
\end{figure}

Figure~\ref{Fig:vcmMM} shows the motor activation rate dependence of
$v_{CM}$ of a polymer with $N=32$ monomers at different values of
the bending parameter (persistent length equal to $0$, $5$, $15$ and
$25 \, d_0$).  We observe that the polymer velocity follows
Eq.~(\ref{mmeq}) as a MM law moderated by the contribution
of the constant force $F_c$ and the effect of the interaction of the
polymer with the membrane walls $F_w$. We can fit our results to
 \be
  v_{CM} \simeq \frac{v_{HR}}{(1+\nu_0/\nu)} + v_{LR}
  \label{eqv}
 \ee
with $v_{HR}$ the high rate limit, and $v_{LR}$ the low rate limit of the system.
The right inset of Fig.~\ref{Fig:vcmMM} confirms the expected number of monomers inside the motor $n_{\rm M} \simeq 5$, value that is observed in all our simulations.

The velocity shows a clear MM-like curve, being dominated by the motor force contributions. Regarding the bending dependence in the velocity,
as anticipated, bending effects come from the $F_w$ term.  In general, to take into account the $F_w$ contribution at different bending, we can write $v_{HR}(k_b) = F_w(\infty,k_b)/N + n_M (F_c+F_M) / N$ and $v_{LR}(k_b) = F_w(0,k_b)/N + n_M F_c / N$. Then, equation~({\ref{eqv}}) allows for a direct experimental fit once the velocity at both high and low ATP concentration are measured at every $k_b$ value.

The force $F_w$ given by the reaction of the walls during the translocation is not easy to evaluate directly but can be computed using Eq.(\ref{mmeq}). This force term, shown in the left inset of Fig.~\ref{Fig:vcmMM} for different bending parameters, slows down the translocation process and appears to be more intense for stiffer chains.
The physical origin of this contribution can be ascribed to the interactions of the polymer with the membrane walls in a direction opposed to the movement. In fact, in the right-hand side of Eq.~\ref{mmeq}, the firsts two terms weakly depend on $k_b$ (visible also in the right inset of Fig.~\ref{Fig:vcmMM}, where a very weak dependence on $k_b$ is present in the number of monomers inside the pore $n_M$). Thus, the main contribution with $k_b$ comes from the $F_w$ term, as far as this model can devise.
A different approach has been followed by Adhikari and Bhattacharya \cite{Adhikari-scaling-2013}.

\subsection*{Translocation time}

\begin{figure}[b]
\centering
\includegraphics[angle=-90, width=8.5cm]{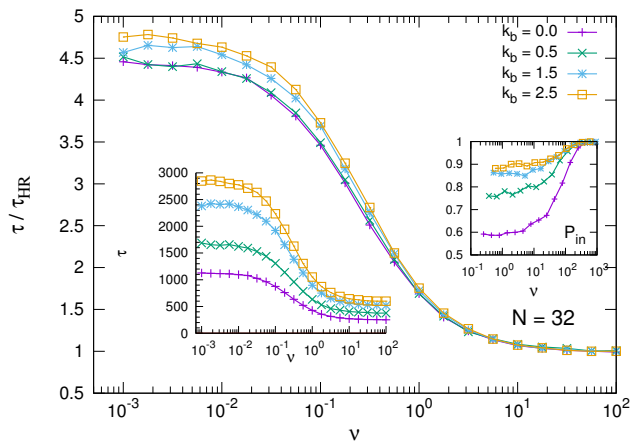}
\caption{Mean translocation time $\tau$ normalized with its high rate limit $\tau_{HR}$ value, for different values of the bending parameter $k_b$, with $k_b = 0.0, 0.5, 1.5, 2.5$, and $N=32$. Right inset: translocation probability $P_{in}$. Left inset: Mean translocation time (not normalized). The values at high rate are $\tau_{HR} = 253, 374, 521, 599$ time units, for the four bending values, respectively.}
 \label{Fig:tau_H_NN_BB}
 \end{figure}

Figure~\ref{Fig:tau_H_NN_BB} shows the mean of the TT $\tau$ of the system as a function of the motor activation rate $\nu$ calculated for $N=32$. $\tau$ decreases monotonously as $\nu$ increases and reaches a limit value for large enough values of $\nu$, when the waiting time goes to zero and the motor stays active most of the time. The results are presented for four different bending values and the curves are normalised with respect the high rate value $\tau_{HR}$ in each case.

 \begin{figure}[tp]
\centering
\includegraphics[angle=-90, width=8.5cm]{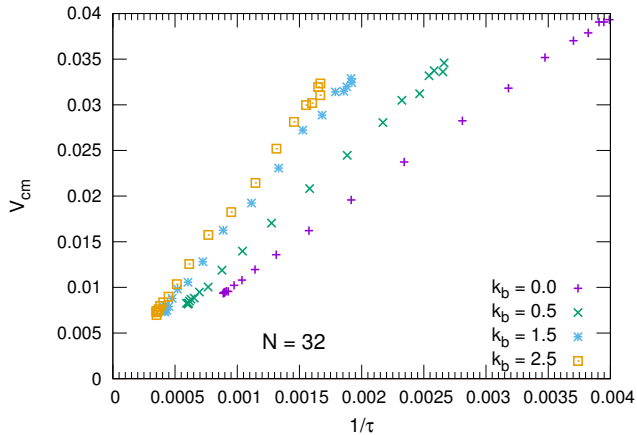}
\caption{Velocity as a function of the $1/\tau$. The linearity of the curves indicates that the relation $v \approx 1/\tau$ holds for a large range of forces here used.}
 \label{Fw}
 \end{figure}
Similar behaviour is observed at other studied
lengths, namely $N=16, 32, 64, 96,$ and $128$. In all cases we find
that $\tau$ increases with $k_b$. In addition, the
translocation probability $P_{in}$ is almost $1$ for all rates at high $k_b$, \emph{i.e.} almost all the simulated events translocate; but $P_{in}$ is lower than 1 in a wide moderate-to-low activation rate region, reaching the value $P_{in}=0.6$ for $k_b=0$. Note that the backward force required to prevent the polymer to translocate has to be stronger for rigid chains than for low $k_b$ values, since the escape from the pore needs a larger displacement of the center of mass of the system to leave the motor, and so, a higher work is required to pull more rigid chains.

Regarding the relation between $v_{CM}$ and $\tau$, it is observed that $v_{CM} \propto 1/\tau$, although with a slope which depends on $k_b$ (see Fig.~\ref{Fw}). The dependence with the bending appears more evident in the $\tau$ curves than in the velocity ones. In fact, as said, more rigid polymers have to displace a longer distance than softer ones to translocate, and they do it at a slower velocity (see Fig.~\ref{Fig:vcmMM}).

\begin{figure}[bp]
\centering
\includegraphics[angle=-90, width=8.5cm]{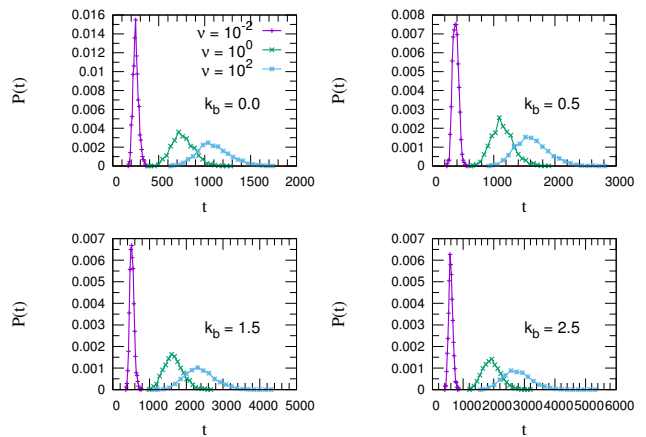}
\caption{Translocation time distribution $P(t)$ for $k_b = 0.0, 0.5,
  1.5, 2.5$ with $N=32$.}
 \label{Pt}
 \end{figure}

Figure~\ref{Pt} shows the TT probability distribution for three values of the activation rates ($\nu = 10^{-2}, 10^{0}, 10^{2}$). For each of the four $k_b$ values we can notice a narrower distribution at higher rates ($\nu=10^{2}$) with a lower mean TT value, with respect to a wider distribution at lower rates ($\nu=10^{-2}$) having a higher mean TT value. All the distributions are almost symmetrical, and this feature can explain the almost linear relationship reported in Fig.~\ref{Fw}.

\subsection*{Translocation time and scaling \label{ScalingSection}}

An interesting issue in this study concerns the scaling properties of the translocation time of the polymer with its size. Below we will propose and check against our numerical results a very simple relation
\be
   \tau = b N (N-n)^{\alpha}
   \label{eq:sc}
\ee
where $\alpha$ is the scaling exponent for the gyration radius of the polymer, $n=n_M+1$  and $b$ the \emph{only} free parameter of the proposed fitting function. We will see that in spite of its simplicity the proposed equation gives account of the observed scaling behaviour of the system in a wide range of rate values.

To depict a simple intuitive model, let us first consider the polymer translocating in a straight line. The translocation length is then $L_t = d_0(N-1-n_M)$, where $n_M$ is the number of monomers remaining
inside the pore, and $d_0$ is the inter-monomer distance. Assuming
that $v_{CM} \propto 1/N$ (Eq.~{\ref{mmeq}) we get $\tau_{exp} \approx L_t/v_{CM} \propto N(N-n), \label{tau-scaling}$ where $n = 1+n_{\rm M} = 6$. For large polymers, the scaling follows the law $\tau \propto N^2$ (see~\cite{deGennes}). The above relationship strictly holds for one-dimensional polymers, where no spatial contributions are present to affect the mean translocation velocity.

The three-dimensional stochastic movement of the polymer outside the
channel gives rise to a general reduction of the translocation time,
which depends on the polymer and environment properties (bending,
exclusion volume, temperature).

\begin{figure}[t]
\centering
\includegraphics[angle=-90, width=8.5cm]{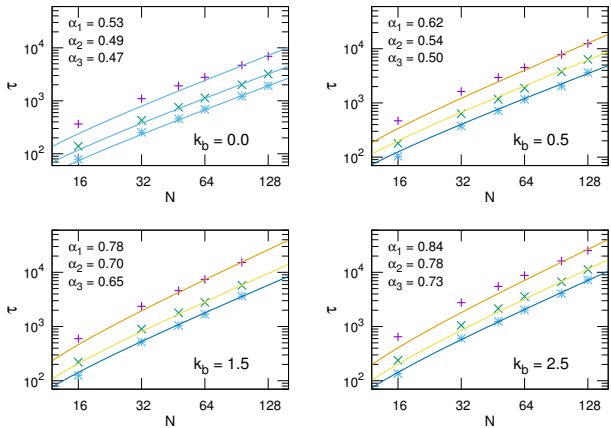}
\caption{$\tau$ at different $N$ values. Each panel corresponds to a different $k_b$ and shows three rate values: $\nu_1=10^{-2}$ (upper
  curves), $\nu_2=1$ (middle curves), $\nu_3=10^{2}$ (lower
  curves). The full lines stand for $\tau=bN(N-6)^{\alpha}$ with $b$ the unique fit parameter.}
 \label{Fig:scaling}
 \end{figure}

\begin{figure*}[tb]
\centering
\includegraphics[angle=-90, width=17cm]{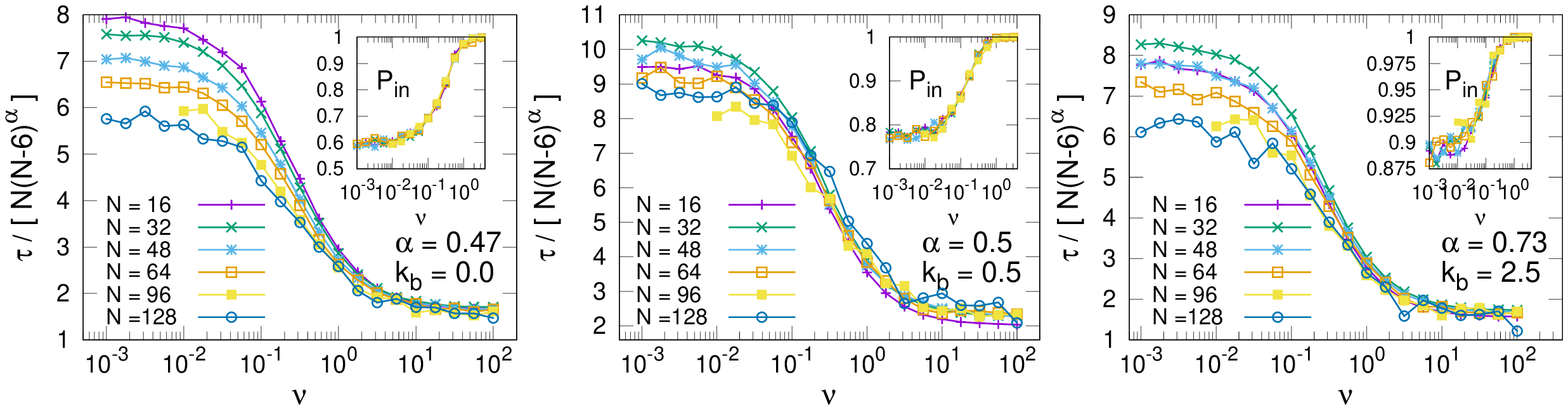}
\caption{Scaled translocation times at three values of the polymer
  bending parameter $k_b=0.0, 0.5$ and $2.5$. The inset shows the
  translocation probability $P_{in}$ for the different polymer
  lengths.}
 \label{Fig:s-B05-n}
 \end{figure*}

In our simple approach, we can assume that the translocation length
changes because of the conformational shape adopted by the polymer at
both sides of the pore. Since the size of the polymer is given by its
gyration radius $R_g$ we can roughly assume $L_t \approx
R_g+(n_M-1)d_0 \simeq R_g$ with $R_g \propto (N-n)^\alpha$.  Thus
$\tau \approx L_t/v_{CM} \propto N (N-n)^\alpha$.

For large polymer this expression becomes $ \tau \propto
N^{1+\alpha}$, in agreement with the time scaling proposed by Kantor
and Kardar \cite{Kardar2004} $\tau \propto N^{1+\nu_F}$, with $\nu_F$
the so-called Flory coefficient.  For a more refined derivation of
this case see also Ref.~\cite{Rowghanian2011} In Flory theory, $\nu_F
= 0.5$ for an ideal chain in 3D (without interaction between no
consecutive monomers), $\nu_F \simeq 0.6$ if we consider the excluded
volume contribution, and $\nu_F = 1$ for a rod (rigid chain)
\cite{Grosberg,luo,Binder2013,Ikonen-scaling-PRE2012,Ikonen-scaling-EPL2013}.

We have computed (see Appendix) the gyration radius of our system at
different bending values and frequencies. We observe it increases with $k_b$, being $\alpha \simeq 0.5$ for $k_b=0$ and tending to $\alpha=1$ as we increase $k_b$. Regarding its variation with the rate $\nu$, a weak dependence is observed, with $\alpha$ decreasing when the rate increases. Values are given inside the related figure.

In figure~\ref{Fig:scaling} we compare our theoretical prediction,
$\tau \propto N(N-6)^\alpha$, with $\alpha$ obtained from the $R_g$
analysis, against our numerical observation for $\tau$. We observe
excellent agreement in all the cases, in spite of the different
approximations employed in the derivation and the moderate to small
size of the chains studied.

Figure~\ref{Fig:s-B05-n} shows the scaled $\tau(\nu)$ curves for different values of $N$ at three bending values, $k_b=0.0, 0.5$ and $2.5$. Here
we use the same scaling exponent $\alpha$ for all the rates (we choose the high rate result). As expected, the scaling is excellent at high $\nu$, while the data dispersion at lower rates indicates that the exponent $\alpha$ depends also on the value of the rate, as explained above. It is worth to note, see the insets, that the translocation probability does not change with $N$, being mostly a function of the bending parameter.

\subsection*{Persistence length analysis}

Persistence length is the basic parameter measuring the stiffness of a polymer. For the continuous WLC model the persistence length is given
by $L_p=k_b/k_BT$. Here we use a discrete version, so deviations from the continuous model may be found for small $N$ values. In order to study the translocation behavior of the polymer at different values of $L_p$ we have performed simulations for different values of the bending from $k_b=0$ (freely-jointed chain model) to high $k_b$ values (extensible rigid chain). The results have been already presented and commented in previous figures and sections of this article.

In Fig.~\ref{Fig:tau_H_N32_BB_Nu} we plot both the translocation time $\tau$ and the center of mass velocity $v_{CM}$ as a function of $L_p$. We show the results for the $N=32$ polymer and at low, medium and high activation rates ($\nu= 10^{-2}, 1, 10^{2}$). We also show the constant force result corresponding to the $\nu \to \infty$ limit. As expected, a saturating behaviour of $\tau$ for increasing values of $L_p$ is observed. The inset shows a similar saturating behaviour for the velocity. The curves for $\nu =\infty$ and $\nu=10^2$ completely overlap as at this latter value the motor is almost continuously actuating.

We have seen that the observables of the system, $\tau$ and $v_{CM}$
depends on $\nu,k_b$ and $N$. However, given the definition of
$L_p=k_b/k_BT$, it is possible to define a new parameter $N/L_p$
giving account of the number of effective segments of the system. The
question now is if it is possible to write either $\tau$ or $v_{CM}$ as a function of $\nu$ and $N/L_p$, thus reducing the number of parameters of the system from 3 to 2. We have investigate this issue and the answer is negative. For a given activation rate, it is observed that simulations data at different values of $N$ and $k_b$ do not collapse to a single curve when plotted as a function of $N/L_p$.

\begin{figure}[]
\centering
\includegraphics[angle=-90, width=8.5cm]{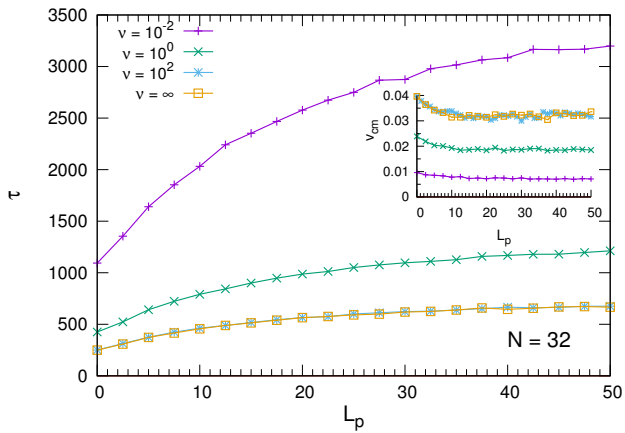}
\caption{Translocation time and center of mass velocity (inset) as a
  function of the persistence length $L_p$ for the cases of constant
  applied force ($\nu=\infty$ in the label) and at different values of   the motor activation rate $\nu = 10^{-2}, 1, 10^{2}$.}
 \label{Fig:tau_H_N32_BB_Nu}
 \end{figure}

\begin{figure*}[t]
\centering
\includegraphics[angle=-90, width=17cm]{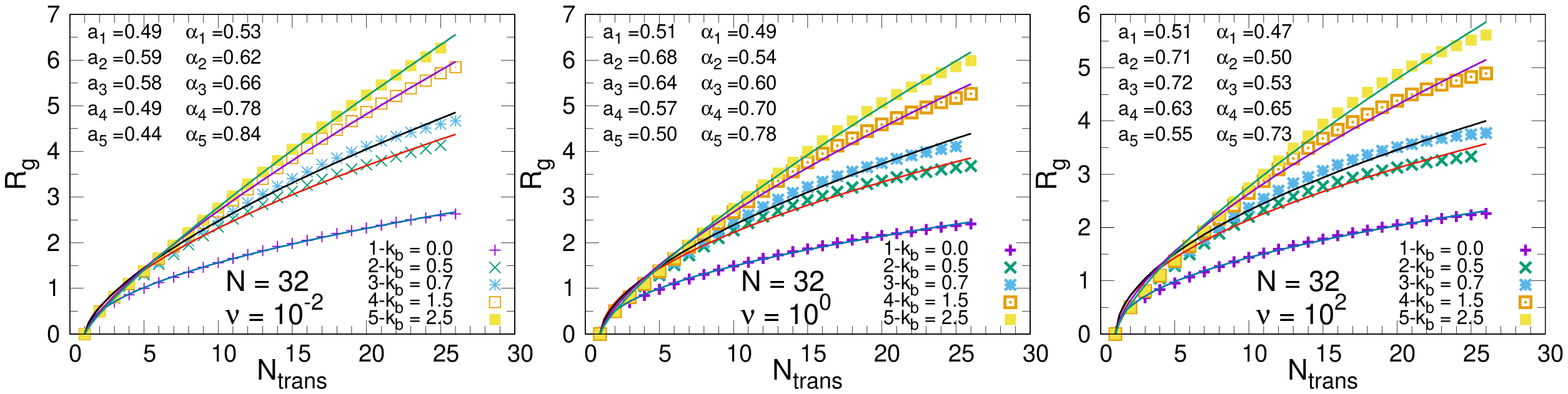}
\caption{Polymer gyration radius as a function of the number of
  monomers reaching the \emph{trans} region, $N_{\rm trans}$, on the
  translocation of $N=32$ polymers. Each point results of an averaging  over the $N_{exp}$ realisations. The curves are fitted to $f(N_{\rm
    trans})=a(N_{\rm trans}-1)^{\alpha}$.}
 \label{Fig:radioGiro_nu}
 \end{figure*}

\section*{Summary and Conclusions}

We have studied the translocation of a polymer chain through a
repulsive, uniform pore membrane in a thermal fluctuating
environment. The polymer is pulled by a time dependent force modelled
as a dichotomous stochastic motor fuelled by the hydrolysis of ATP molecules which bind to the motor at a Poisson rate and
activates a mechanical work during a fixed time lapse.

The objectives of our work are to study the specificity of the
pull force used inside the pore, and to investigate the
translocation time dependence on the flexibility and length of the
polymer chain. We propose a scaling behaviour of the translocation
times with the polymer length according to a power law of the type
$N^{1+\alpha}$, with $\alpha$ weakly dependent on the activation rate
of the motor.

This work is the natural extension of our previous
studies~\cite{ajf-sin,ajf-rtn,ajf-damn,ajf-sin-rtn-3D} to a more
realistic and biologically interesting system, accounting for motion
in the 3-D domain, polymer-membrane interactions, and ATP based
driving motor.  Dynamics has been studied for different motor
actuation rates, polymer sizes, and rigidities.

The nature of the ATP-based pulling gives rise to monotonic
translocation time as a function of the activation rate of the motor.
The mean velocity of the polymer is found to have a specific and very
clear behaviour, \emph{i.e.} a MM dependence with the activation rate of the motor, which directly depends on the ATP molecule concentration in the surroundings of the machine.  The MM velocity for an ATP-driven machine has been extensively measured in many experiments with
different types of motors. Examples are the transport velocity of a
stepping motor like kinesins \cite{1994Svoboda,2000Schnitzer} (with or without load forces) as well as the packaging velocity of the DNA of
the bacteriophage pulled inside the virus capside by its motor
complex\cite{Bust05}.

We show here that such MM-like behaviour arises in a natural way in
the introduced model. Specifically, the translocation velocity depends on
the motor activity which drives the polymer for a certain fixed time, and stays inactive during a certain Poisson distributed waiting
time. The {\it average working time} of each molecular motors using
ATP in these conditions, results in a MM kinetics.

Beyond its biological interest, the study of polymer translocation through active nanopores reveals its importance in understanding the
translocation aspects of long molecules assisted by motors at small scales, and represents a starting point for the construction from scratch of nanotechnological objects. In this regard, without any purpose to reproduce in detail some specific experiment, this work is an attempt to model the dynamics at the nanoscale with a very comfortable approach under the realistic constraints imposed by the ATP energy supply of a typical molecular motor. Further improvement of this kind of description can be devoted to a more accurate modelling of the pore channel in both: the structure of the very pore together with its interaction with the passing through molecule, and a more complex characterisation of the polymer by enlarging its inner degrees of freedom such as the diversification of the monomers properties according to the different molecular components of a real chain.

\appendix*
\section*{Gyration radius}

This appendix is devoted to provide details on the numerical
calculation of the exponent $\alpha$ introduced in our scaling
analysis of the polymer translocation time. This is based in the
computation of a {\em dynamical} gyration radius as follows: given a
polymer with $N$ monomers we define $N_{\rm trans}$ as the number of
monomers in the \emph{trans} side of the system (evaluated at the time
instant a monomer reaches the side). At this time we compute the
instantaneous gyration radius of the system, and average it over
$N_{exp}$ translocation realisation:
\be
  R_G = \left \langle \sqrt{\frac{1}{N_{\rm trans}}\sum_{i=1}^{N_{\rm trans}} (\mathbf{r}_{\rm i}-\mathbf{r}_{CM})^2} \right \rangle_{N_{exp}}.
\ee
This number behaves as $R_g= a(N_{\rm trans}-1)^{\alpha}$.

Fig.~\ref{Fig:radioGiro_nu} shows the behaviour of this gyration
radius as a function of the number of monomers which have crossed the
membrane, for the case os a $N=32$ beads polymer and at different values
of $\nu$ and $k_b$. Fitting is always good and the exponents (given in
the figure), decrease a bit increasing the rate, but depends more
strongly on the stiffness of the system, $k_b$. As expected the
gyration radius (and the scaling exponent) increases with $k_b$.  The
obtained exponents were used in the scaling study of the polymer
translocation time as discussed in Section~\ref{ScalingSection}.

We have to mention that differently from the usual definition of the
gyration radius, which makes sense in thermodynamic equilibrium
conditions, here we calculate a somehow {\em dynamical} gyration
radius. Evidently, the calculation of any magnitude along the
translocation process, constitute an out of equilibrium measure,
though the slower is the translocation dynamics, the better is the
approximation to a thermodynamic equilibrium (thus best fitting is
obtained at small activation rates). In this sense is not surprising
that the gyration radius is a function of both the activation rate of
the motor (and so of the translocation velocity), and, of course,
strongly dependent of the stiffness of the chain. That is why we can
observe in Fig.~\ref{Fig:radioGiro_nu} that the radius slowly
increases by decreasing the activation rates.

\section*{Acknowledgments}
This work is supported by the Spanish project Mineco No.~FIS2014-55867, co-financed by FEDER funds. We also thank the support of the Arag\'on Government and Fondo Social Europeo to FENOL group.

\section*{Author contributions statement}
A.F. conducted the simulations, A.F. and F.F. designed the project. A.F., F.F. and J.M. analyzed the results. All authors reviewed the manuscript.

\end{document}